\documentclass{phbauth}
\usepackage{graphicx}
\newcommand{\text}[1]{{\mathrm{#1}}}
\renewcommand{\vec}{\mathbf}

\begin{document}
\begin{frontmatter}
\title{Hole-hole correlation effects on magnetic properties
of Mn$_x$III$_{1-x}$V diluted magnetic semiconductors}

\author[ind,cz,aus]{T. Jungwirth\thanksref{corresp}},
\author[ind,aus]{Byounghak Lee},
\author[ind,aus]{A.H. MacDonald}

\address[ind]{Department of Physics, Indiana University,
              Bloomington, IN 47405, USA}
\address[cz]{Institute of Physics ASCR, Cukrovarnick\'a 10, 
        162 00 Prague 6, Czech Republic}
\address[aus]{Department of Physics, The University of Texas at Austin,
              Austin, TX 78712, USA}

\thanks[corresp]{Corresponding author: fax: +1-812-855-5533;
E-mail address: jungw@gibbs.physics.indiana.edu}

\begin{abstract}
The mean-field theory represents a useful
starting point for studying carrier-induced ferromagnetism
in Mn$_x$III$_{1-x}$V diluted magnetic semiconductors.
A detail description of these systems
requires to include correlations in the many-body hole system.
We discuss the effects of correlations among itinerant carriers
on magnetic properties of bulk Mn$_x$III$_{1-x}$V and magnetic
semiconductor quantum wells. Presented results were obtained using
parabolic band approximation and we also derive a many-body perturbation
technique that allows to account for
hole-hole correlations in realistic semiconductor valence bands.
\end{abstract}
\end{frontmatter}

\section{Introduction}

Recent
advances \cite{ohno:apl96,ohno:jmmm99,dietl:sci00} in
the fabrication and control of Mn$_x$III$_{1-x}$V
diluted magnetic semiconductors (DMS)
have
opened up a broad and relatively unexplored frontier for both basic and applied
research. We have developed a mean-field theory of free-carrier induced
ferromagnetism in DMS which is intended to be useful for 
bulk materials as well as for any spatially inhomogeneous systems. 
Special emphasis is placed on interaction effects in the itinerant 
many-body system whose inclusion is required for a detailed understanding
of magnetic properties of these materials. In Section~2 we derive the
mean-field theory equations. Correlation effects on magnetic properties
of bulk DMS and Mn-doped semiconductor
quantum wells are studied in Sections
3 and 4, respectively, using a parabolic band model. In Section 5
we present a many-body perturbation theory that accounts for
hole-hole interactions in realistic semiconductor valence bands and
evaluate the exchange enhancement of the ferromagnetic critical 
temperature for typical sample parameters.

\section{Mean-field theory}

Our theory is based
on an envelope function description of the valence band electrons
and a spin representation of their kinetic-exchange
interaction \cite{dmsreviews} with
d-electrons on the $S=5/2$ Mn$^{++}$ ions:
\begin{equation}
{H} = {H}_m + {H}_f + J_{pd} \sum_{i,I} {\vec S_I}
\cdot {\vec s}_i \delta({\vec r}_i - {\vec R}_I),
\label{coupling}
\end{equation}
where $i$ labels a free carrier, $I$ labels a magnetic
ion and the exchange interaction is parametrized by a coupling
constant $J_{pd}$.
In Eq.~(\ref{coupling}) ${H}_m$ is the Hamiltonian of the
magnetic ions, ${H}_f$ is the six-band Luttinger Hamiltonian
for free carriers in the valence band,  $\vec S$ is the magnetic ion  spin and
$\vec s$ is the electron-spin operator projected onto the $j=3/2$ and 1/2
valence band manifold of the Luttinger Hamiltonian.

In the absence of external fields, the mean polarization of a
magnetic ion is given by \cite{usprb}
\begin{equation}
\langle m \rangle_I = S B_{S}\big(J_{pd} S \big[n_{\uparrow}(\vec R_I)
-n_{\downarrow}(\vec R_I)\big]/2 k_B T\big)\; ,
\label{mI}
\end{equation}
where $B_S(x)$ is the Brillouin function,
\begin{equation}
B_S(x) 
\approx {S+1 \over 3S}x, \: x \ll
1\, .
\end{equation}
The electron spin-densities $n_{\sigma}(\vec r)$ are
determined by solving the Schr\"{o}dinger equation
for electrons which experience an electrostatic potential,
$v_{es}(\vec r)$, a spin-dependent kinetic-exchange potential,
\begin{equation}
h_{pd}(\vec r) = J_{pd} \sum_I \delta (\vec r - \vec R_I) \langle m
\rangle_I\; ,
\label{exchangefield}
\end{equation} 
and a local-spin-density-approximation (LSDA) exchange-correlation potential,
$v_{xc}(\vec r)$, on which we comment further
below.
The kinetic-exchange potential is non-zero only in the ferromagnetic state
and we assume that the magnetic ions are randomly distributed
and dense on a scale set by the free carrier Fermi wavevector 
and that their density, 
$c(\vec r)$, can be  controlled.
This allows us to take a continuum limit where 
\begin{equation}
h_{pd}(\vec r) = J_{pd} \  c(\vec r)\ \langle m \rangle(\vec r).
\label{continuum}
\end{equation}

\section{Homogeneous DMS}

\begin{figure}
\centerline{\includegraphics[width=8cm]{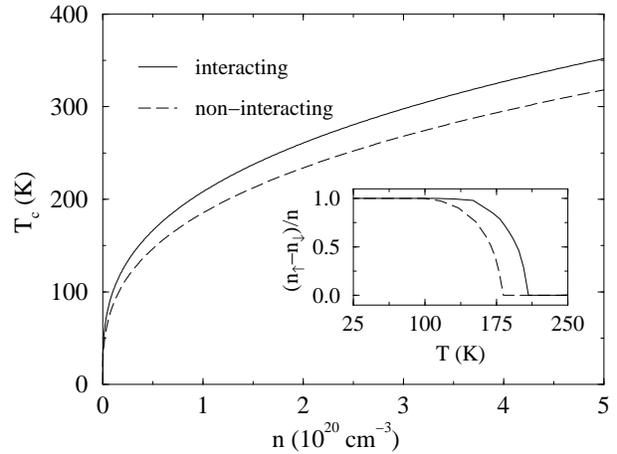}}
\caption{Curie-Weiss temperature as a function of hole density calculated
including (solid line) and neglecting (dashed line) hole 
exchange-correlation potential. Inset: relative spin polarization
of holes as a function of temperature for $n=10^{20}$~cm$^{-3}$.}
\label{tcb}
\end{figure}

For homogeneous systems with randomly
distributed localized spins the mean-field equations can be
solved analytically. The hole spin-density and the kinetic-exchange
potential are related as
\begin{equation}
(n_{\uparrow}-n_{\downarrow})/2=
\frac{\chi_f}{(g^* \mu_B)^2}h_{pd}\; ,
\label{bulk}
\end{equation}
where $\chi_f$ is the interacting hole magnetic susceptibility.
The Curie-Weiss temperature, obtained from Eqs.~(\ref{mI})
- (\ref{bulk}), is given by
\begin{equation}
k_B T_c = \frac{c S (S+1)}{3} \frac{J_{pd}^2\chi_f}{(g \mu_B)^2}
\; .
\label{tc}
\end{equation}
In Fig.~\ref{tcb} we have
plotted the ferromagnetic transition temperature as a function
of the total carrier density, $n$, predicted by
this expression for $p$-type Mn$_x$Ga$_{1-x}$As
with  $J_{pd}=0.15$ eV~nm$^3$
$c=10^{21}$ cm$^{-3}$ and for valence bands approximated
by a single parabolic band with effective hole  mass $m^*=0.5m_e$.
When exchange and correlation effects are neglected
and $\chi_f$ is replaced by its zero-temperature value,
$T_c$ is  proportional to $n^{1/3}$.
Including the parabolic band exchange-correlation potential
\cite{vosko}
enhances $T_c$ by $\approx 20\%$ for
typical hole densities $\sim 10^{20}$~cm$^{-3}$.

\section{DMS quantum wells}
In the parabolic band approximation,
the spin-densities in Mn$_x$Ga$_{1-x}$As quantum wells are determined 
by solving the Schr\"{o}dinger equation 
\begin{eqnarray}
& &\bigg[ - \frac{ 1}{2 m^*_{\parallel}}\big(\frac{\partial^2}{\partial x^2}
+\frac{\partial^2}{\partial y^2}\big)-
\frac{ 1}{2 m^*_{z}}\frac{\partial^2}{\partial z^2}
+ v_{es}(z)  \nonumber \\
& &+v_{xc,\sigma}(z)
 - \frac{\sigma}{2} h_{pd}(z) \bigg] \psi_{k,\sigma}(\vec r)
 = \epsilon_{k,\sigma} \psi_{k,\sigma}(\vec r)\; ;
\nonumber
\label{kseqs}
\end{eqnarray}
\begin{equation}
n_{\sigma}(z) = \sum_k f(\epsilon_{k,\sigma}) |\psi_{k,\sigma}(z)
|^2\; ,
\label{densities}
\end{equation}
where the in-plane effective mass
$m^*_{\parallel} \approx
0.11m_0$ and the out-of-plane mass $m^*_{z} \approx
0.38m_0$. The electrostatic potential varies along the growth
($z$) direction and includes band offset and ionized impurity
contributions. The LSDA equation for the exchange-correlation
potential reads
\begin{equation}
v_{xc,\sigma}(z) = \frac{d [n
\epsilon_{xc}(n_{\uparrow},n_{\downarrow})]}{d n_{\sigma}}
\big|_{n_{\sigma}= n_{\sigma}(z)}\; ,
\label{xcpot}
\end{equation}
where $\epsilon_{xc}(n_{\uparrow},n_{\downarrow})$
is the exchange and correlation energy per particle of a spatially
uniform  system \cite{vosko}.

\begin{figure}
\centerline{\includegraphics[width=8cm]{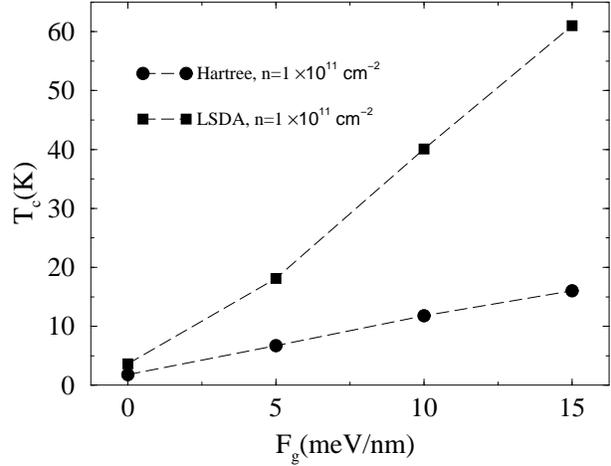}}
\caption{
Dependence of $T_c$ on bias voltage applied across  the
$w=10$~nm wide quantum well partially occupied
by magnetic ions over a distance of 3~nm near one edge of the quantum well.
Circles (squares) represent results of the full numerical self-consistent
calculations without (with) the exchange-correlation potential.
}
\label{tcfg}
\end{figure}

If hole correlations and subband mixing are neglected the following
analytic expression can be found for $T_c$:
\begin{equation}
T_c = {S (S+1) \over 12} {J_{pd}^2 \over k_B} 
{m^{\ast}_{\parallel} \over \pi \hbar^2}
\int dz |\varphi(z)|^4 c(z) \:.
\label{tcqw}
\end{equation}
In Eq.~(\ref{tcqw}), the critical temperature increases with
the subband wavefunction, $\varphi(z)$, and magnetic impurities
overlap, is proportional to $J_{pd}^2$ and $m^*_{\parallel}$.
Similar $T_c$-equation was derived \cite{haury:prl97}
using the RKKY theory and its
predictions are in agreement with the measured carrier-induced
ferromagnetism in II-VI quantum wells
\cite{haury:prl97,kossacki:physicae00}. Here we emphasize the role
of hole-hole interactions and Mn doping profile on magnetic
properties of biased III-V DMS quantum wells. 
A remarkable feature of quasi-2D ferromagnetic
DMS is the possibility of tuning $T_c$ through a wide range
in situ, by the application of a gate voltage.  In
Fig.~\ref{tcfg} we illustrate this for the case of a
$w=10$~nm  quantum well with magnetic ions covering
only a $d=3$~nm portion near one edge.
Even for such a narrow quantum
well, the critical temperature can be varied over an order of
magnitude by applying a bias voltage which draws  electrons into
the magnetic ion region. Hole-hole interactions substantially
increase the critical temperature, as also shown in Fig.~\ref{tcfg}

%
%\begin{figure}
%\centerline{\includegraphics[width=8cm]{tcfg.eps}}
%\caption{Dependence of $T_c$ on bias voltage applied across  the
%$w=10$~nm wide quantum well partially occupied
%by magnetic ions over a distance of 3~nm near one edge of the quantum well.
%Circles (squares) represent results of the full numerical self-consistent
%calculations without (with) the exchange-correlation potential.}
%\label{tcfg}
%\end{figure}
%One remarkable feature of quasi-2D ferromagnetic
%DMS's is the possibility of tuning $T_c$ through a wide range
%in situ, by the application of a gate voltage.  In
%Fig.~\ref{tcfg} we illustrate this for the case of a
%$w=10$~nm  quantum well with magnetic ions covering
%only a $d=3$~nm portion near one edge.
%Even for such a narrow quantum
%well, the critical temperature can be varied over an order of
%magnitude by applying a bias voltage which draws  electrons into
%the magnetic ion region. Fig.~\ref{tcfg} also shows a dramatic
%enhancement of $T_c$ in biased quantum wells due to exchange
%and correlation effects in the hole system. 

\section{Exchange and correlations in realistic valence bands}
A semiempirical six band model is used for quantitative description
of electronic states
in cubic semiconductors near the top of the valence band. For
systems with spin-orbit coupling the Kohn-Luttinger Hamiltonian 
\cite{luttinger:pr55} is
derived in the basis of total angular momentum eigenstates with $j=3/2$
and 1/2, and $m_j=-j,-j+1,...,j$. The many-body interaction Hamiltonian
in this basis can be written as
\begin{equation}
H_I=\sum_{\stackrel{\scriptstyle n_1,n_2,}{\vec{k}_1,\vec{k}_2,\vec{q}}}
a^{\dagger}_{n_2,\vec{k}_2-\vec{q}}
a^{\dagger}_{n_1,\vec{k}_1+\vec{q}}
a_{n_1,\vec{k}_1}
a_{n_2,\vec{k}_2} V(\vec{q}),
\end{equation}
where $V(\vec{q})$ is the Fourier transform of the Coulomb
interaction. Using the following notation for the transformation
between creation and annihilation operators for the total angular
momentum eigenstates and Kohn-Luttinger Hamiltonian eigenstates:
\begin{equation}
a^{\dagger}_{n,\vec k}
%&=&\sum_{\alpha}\langle\psi_{\alpha,\vec k}|
%\varphi_{n,\vec k}\rangle\, a^{\dagger}_{\alpha,\vec k}
%&\equiv&
=
\sum_{\alpha}\overline{z}_{n,\alpha,\vec{k}}\, a^{\dagger}_{\alpha,\vec k}
,\;
%\nonumber \\
a_{n,\vec k}
%&=&\sum_{\alpha}\langle\varphi_{n,\vec k}|
%\psi_{\alpha,\vec k}\rangle\, a_{\alpha,\vec k}
%&\equiv&
=
\sum_{\alpha}z_{n,\alpha,\vec{k}}\, a_{\alpha,\vec k}
\end{equation}
the exchange contribution to the hole total energy has a form
\begin{eqnarray}
E_x&=\frac12
&\sum_{\stackrel{\scriptstyle \alpha_1,\alpha_2,}{\vec{k},\vec{q}}}
n_{\alpha_1,\vec{k}}n_{\alpha_2,\vec{k}+\vec{q}}
\nonumber \\
&\times&\left|\sum_nz_{n,\alpha_2,\vec{k}+\vec{q}}
\overline{z}_{n,\alpha_1,\vec{k}}\right|^2 V(\vec{q})\; ,
\label{ex}
\end{eqnarray}
where $n_{\alpha,\vec k}$ is the Fermi distribution function of
Kohn-Luttinger eigenstates.

In the random phase approximation (RPA), the correlation energy is
related to the wavevector and frequency dependent dielectric function
as \cite{fw}
\begin{eqnarray}
& &E_c=\frac12 \hbar\sum_{\vec q}\int\frac{d\omega}{2\pi}
\bigg\{\tan^{-1}\left[\frac{{\mathrm Im}\epsilon_{RPA}(\vec q,\omega)}
{{\mathrm Re}\epsilon_{RPA}(\vec q,\omega)}\right] \nonumber \\
&-&
{\mathrm Im}\epsilon_{RPA}(\vec q,\omega)
\bigg\}\nonumber \\
& &\epsilon_{RPA}(\vec q,\omega)=1-V(\vec q)D^0(\vec q,\omega)\; ,
\label{corr}
\end{eqnarray}
where $D^0(\vec q,\omega)$ is the semiconductor valence band
density-density response function:
\begin{eqnarray}
& &D^0(\vec q,\omega)=
\sum_{\stackrel{\scriptstyle \alpha_1,\alpha_2,}{\vec{k}}}
n_{\alpha_1,\vec{k}}(1-n_{\alpha_2,\vec{k}+\vec{q}})
\nonumber \\
&\times&
\left|\sum_nz_{n,\alpha_2,\vec{k}+\vec{q}}
\overline{z}_{n,\alpha_1,\vec{k}}\right|^2
\nonumber \\
&\times&
\bigg[\frac{1}{\hbar\omega+(\xi_{\alpha_1,\vec{k}}-
\xi_{\alpha_2,\vec{k}+\vec{q}})+i\eta}\nonumber \\
&-&
\frac{1}{\hbar\omega-(\xi_{\alpha_1,\vec{k}}-
\xi_{\alpha_2,\vec{k}+\vec{q}})-i\eta}\bigg]\; ,
\end{eqnarray}
and $\xi_{\alpha,\vec{k}}$ are the Kohn-Luttinger eigenenergies.
The Curie-Weiss temperature for realistic valence bands is obtained
from Eq.~(\ref{tc}) with the magnetic susceptibility given by
\begin{equation}
\frac{\chi_f}{(g\mu_B)^2}=\frac{d^2(E_T/\Omega)}{dh^2}\; ,
\end{equation}
where the itinerant system total energy per volume, 
$E_T/\Omega$, has a kinetic (band)
contribution and the exchange and correlation contributions given
by Eqs.~(\ref{ex}) and (\ref{corr}).

For the current experimental values of the hole density
$n=3.5\times10^{20}$~cm$^{-3}$, Mn concentration 
$c=1\times10^{21}$~cm$^{-3}$, and kinetic exchange constant
$J_{pd}=55$~meV~nm$^3$, the six-band model gives mean-field critical
temperature for bulk Mn$_x$Ga$_{1-x}$As, $T_c=86$~K when exchange
and correlations are neglected. The exchange and correlation contributions
to the total energy  enhances $T_c$ by a factor
of 1.2. 
Compared to the experimental 
$T_c=110$~K, the mean-field theory accurately predicts  the ferromagnetic
critical temperature of Mn$_x$Ga$_{1-x}$As DMS. 
\section*{Acknowledgments}
We acknowledge helpful interactions with M. Abolfath,
T. Dietl, J. Furdyna,
J. K\"{o}nig, and H. Ohno. The work was performed under
NSF grants DMR-9714055 and DGE-9902579 and Ministry of Education
of the Czech Republic grant OC P5.10.

\end{document}